\documentclass[aps,prd,twocolumn,groupedaddress,
showpacs,
preprintnumbers,amsmath,amssymb]{revtex4}

\usepackage{graphicx,bm}

\allowdisplaybreaks

\begin{document}

\preprint{MAN/HEP/2015/11}

\title{Mass Bounds on Light and Heavy Neutrinos from Radiative MFV Leptogenesis}

\author{Apostolos Pilaftsis}

\author{Daniele Teresi}

\affiliation{Consortium for Fundamental Physics, School of Physics and
  Astronomy, University of Manchester, Manchester M13 9PL, United
  Kingdom.} 

\date{June 29, 2015}

\begin{abstract}

We derive novel  limits on the masses of the  light and heavy Majorana
neutrinos  by requiring  successful leptogenesis  in seesaw  models of
minimal  flavour  violation   (MFV).   Taking  properly  into  account
radiative flavour effects and avoiding  the limitations due to a no-go
theorem on leptonic asymmetries, we find that the mass of the lightest
of  the  observable neutrinos  must  be  smaller than  $\sim~0.05$~eV,
whilst the Majorana scale of  lepton number violation should be higher
than $\sim~10^{12}$~GeV.  The latter  lower bound enables one to probe
the existence of  possible new scales of MFV, up  to energies of $\sim
100$~TeV,  in low-energy experiments,  such as  $\mu \to  e\gamma$ and
$\mu  \to  e$ conversion  in  nuclei.   Possible  realizations of  MFV
leptogenesis in Grand Unified Theories are briefly discussed.

\end{abstract}


\maketitle

The hypothesis of  Minimal Flavour Violation (MFV)~\cite{GIM} provides
an elegant  framework, even  for flavour theories  of new  physics, to
naturally implement the strong constraints from the non-observation of
sizeable  flavour  changing  neutral  currents in  the  quark  sector.
Interestingly  enough, this  MFV  hypothesis can  be  extended to  the
lepton sector as well. In  particular, one may consider seesaw models,
in which all effects  of Lepton Flavour Violation~(LFV), including the
observable   light-neutrino  masses   and   mixing,  originate   {\em
  exclusively}  from  the Yukawa  interactions  to three  right-handed
neutrinos   $N_{R,\alpha}$, with $\alpha \! = \! 1,2,3$, in   a   flavour  basis   with   diagonal
charged-lepton Yukawa  couplings.  Instead, the  right-handed Majorana
mass  matrix  $\bm{M}_N$  takes   on  its  maximally  symmetric  form:
$\bm{M}_N = m_N  {\bf 1}_3$, which means that  $\bm{M}_N$ is invariant
under                                                      $O(3)_{N_R}$
rotations~\cite{Pilaftsis:2005rv,Cirigliano:2005ck}.     Such   seesaw
models allow for  large hierarchies between the seesaw  scale $m_N$ of
Lepton  Number  Violation (LNV)  and  possible  other  scales of  LFV,
e.g.~due  to soft  breaking  of supersymmetry~\cite{Borzumati:1986qx},
giving rise to observable rates for  $\mu \to e \gamma$, $\mu \to eee$
and $\mu \to e$ conversion in nuclei~\cite{Ilakovac:2012sh}.

In addition,  the presence of  heavy Majorana neutrinos  provides, via
the well-established mechanism of leptogenesis~\cite{Fukugita:1986hr},
one of  the most plausible explanations  for the origin  of the Baryon
Asymmetry in the Universe~(BAU). In the MFV seesaw scenario, the exact
mass degeneracy of the three right-handed neutrinos leads to vanishing
leptonic         asymmetries~\cite{Pilaftsis:1997jf}.         However,
Renormalization  Group~(RG)  effects due  to  running  from the  Grand
Unified Theory~(GUT)  scale $\mu_X \approx  2\times 10^{16}~{\rm GeV}$
to a  lower heavy-neutrino scale $m_N$ lift  the $O(3)_{N_R}$ symmetry
in   the  heavy-neutrino   sector,  thus   potentially   offering  the
possibility to explain the BAU via the so-called mechanism of Resonant
Leptogenesis (RL)~\cite{Pilaftsis:1997jf,Dev:2014laa}.

In    this   \emph{radiative}   framework~\cite{GonzalezFelipe:2003fi,
  Cirigliano:2006nu,   Branco:2006hz}  of   MFV,  it   was  originally
argued~\cite{Cirigliano:2006nu} that  successful leptogenesis requires
the LNV seesaw scale $m_N$ to be higher than $10^{12}\, \mathrm{GeV}$,
implying sizeable neutrino Yukawa  couplings larger than $\sim0.1$, so
as to fit the low-energy neutrino data.  The latter would allow one to
test the possible existence of additional new-physics scales mediating
minimal   LFV,   for   a    wide   range   of   energies   from~1   to
100~TeV~\cite{Cirigliano:2006nu},  in low-energy experiments,  such as
the     upgraded     MEG     experiment~\cite{Baldini:2013ke}.      Subsequent     elaborate
studies~\cite{Branco:2006hz}, however,  which include flavour effects,
have  suggested that $m_N$  could be  as low  as $10^6\,\mathrm{GeV}$,
implying neutrino Yukawa couplings that could be as small as $10^{-4}$.  Thus, these
results seem to eliminate any prospects of definitely excluding the presence of new
MFV  scales in  the  near  future.  Most  recently,  however, a  no-go
theorem~\cite{Dev:2015wpa} for vanishing leptonic asymmetries to order
$h^4$  in the  neutrino Yukawa  couplings has  been  proved, rendering
radiative MFV leptogenesis not viable in the low-$m_N$ region.

In this article we will show how this no-go theorem can be circumvented
in   a   MFV   framework,   provided   the   seesaw   scale   $m_N   >
10^{12}\,\mathrm{GeV}$.   The  latter reopens  the  way of  falsifying
possible  new leptonic  MFV  scales up  to energies  of~$\sim100\,{\rm
  TeV}$.  A complete treatment  of flavour effects yields novel limits
on the masses of the light  neutrinos as well.  In particular, we find
that  the  mass  of  the   lightest  neutrino  must  be  smaller  than
$\sim0.05$~eV,  with the  normal hierarchical  light-neutrino spectrum
favoured over the inverted one.

In  the  MFV  seesaw  scenario,  the  flavour-symmetric  part  of  the
Lagrangian $\mathcal L_{\mathrm{sym}}$ at the GUT scale $\mu_X$ reads
\begin{equation}\label{eq:lagr}
\mathcal L_{\mathrm{sym}} \ = \ \mathcal
L^{\mathrm{SM}}_{\mathrm{sym}} \;+\; \frac{1}{2} \overline{N}_{\rm R,
  \alpha}^C  [M_N]^{\alpha \beta}  
  N_{\rm R, \beta} \;+\; {\rm H.c.}\;, 
\end{equation}
where $[M_N]^{\alpha \beta} = m_N \delta^{\alpha \beta}$ and $\mathcal
L^{\mathrm{SM}}_{\mathrm{sym}}$ is the Standard Model (SM) Lagrangian,
without the  neutrino and charged-lepton  Yukawa interactions mediated
by the couplings $h_l^{\phantom l \alpha}$ and $y_{l}^{\phantom l m}$,
respectively. In this work, we use the flavour-covariant formalism developed in~\cite{Dev:2014laa}. Under general unitary flavour transformations $U(3)_L  \times U(3)_{N_R}$, the charged-lepton SM doublets $L_l$ (with $l\!=\!1,2,3$) and the heavy neutrinos $N_{\rm R, \alpha}$ transform as
\begin{subequations}\label{eq:trans_fields}
\begin{align}
  L_l \ &\rightarrow \ L'_l \ = \ V_l^{\phantom l m} \; L_m \;, \\
  L^{l} \ & \equiv\ (L_l)^\dagger \ \rightarrow \ L'^l \ = \ 
  V^l_{\phantom{l} m} \; L^{m} \;, \\
  N_{\rm R, \alpha} \ &\rightarrow \ N'_{\rm R, \alpha} \ = \ 
  U_\alpha^{\phantom \alpha \beta} \;
  N_{\rm R, \beta} \;, \\
  N_{\rm R}^{\phantom{\rm R} \alpha} \ & \equiv \
  (N_{\rm R, \alpha})^\dagger \ \rightarrow \ 
  N_{\rm R}'^{\phantom{\rm R} \alpha} \ = \ 
  U^\alpha_{\phantom{\alpha} \beta} \;
  N_{\rm R}^{\phantom{\rm R} \beta} \;,
  \label{flavtrans_N}
\end{align}
\end{subequations}
with  the  unitary transformation  matrices  $V_l^{\phantom  l m}  \in
U(3)_L$   and    $U_\alpha^{\phantom   \alpha   \beta}   \in
U(3)_{N_R}$. All the other tensors, e.g. the heavy-neutrino Yukawa couplings $h_l^{\phantom l \alpha}$, transform analogously as indicated by the position of the flavour indices.

  The  Lagrangian  $\mathcal{L}_{\mathrm{sym}}$ is  invariant under  the lepton-flavour symmetry  group $G_{\mathrm{LF}} =
U(3)_L  \times U(3)_{e_R} \times  O(3)_{N_R}$, where $e_R$ collectively
denotes  the three right-handed charged  leptons.  According
to  the  MFV hypothesis,  the  flavour  symmetry $G_{\mathrm{LF}}$  is
explicitly  broken only by  the   couplings  $y_{l}^{\phantom  l  m}$  and
$h_l^{\phantom  l \alpha}$,  which occur  in the  Lagrangian $\mathcal
L_{\mathrm{Y}}$ of the leptonic Yukawa sector:
\begin{equation}
\mathcal{L}_{\mathrm{Y}} \  = \  y_{l}^{\phantom l m} \overline L^{l} 
 \Phi \,e_{R,m} \;+\; h_{l}^{\phantom l \alpha} \overline L^{l} 
  \widetilde{\Phi}  N_{\rm R, \alpha} 
   \;+\; {\rm H.c.}\;,
\end{equation}
where  $\widetilde{\Phi}  \equiv  i  \sigma_2  \Phi^*$  is  the  Higgs
hypercharge   conjugate.   

As mentioned above,  the $O(3)$ mass degeneracy of  the heavy-neutrino
spectrum will be lifted by RG  effects when running from the GUT scale
$\mu_X$  to   $m_N$.   By  means  of  the   variables  $t(\mu)  \equiv
\ln(\mu_X/\mu)$  and   $t_N  \equiv  t(m_N)$,  the   RG  equation  for
$\bm{M}_N$ is given by~\cite{Chankowski:1993tx}
\begin{equation}\label{eq:RG}
\frac{d \bm{M}_N}{d t} \ \simeq \ - \, \frac{1}{16 \pi^2} \, \Big[
  \big( \bm{h}^\dag \bm{h} \big) \bm{M}_N \,+\, \bm{M}_N \big(
  \bm{h}^T \bm{h}^* \big)\Big]\;. 
\end{equation}
At leading order in  $\bm{h}^\dag \bm{h}$, the $O(3)$-breaking part of
$\bm{M}_N$ may be evaluated as
\begin{align}
 \bm{\Delta M}_N^{\rm RG} \ &= \ t_N \, \frac{d \bm{M}_N}{d
   t}\bigg|_{t=0} +\; \frac{t_N^2}{2} \, \frac{d^2 \bm{M}_N}{d
   t^2}\bigg|_{t=0} + \; \ldots \notag\\ &\simeq \ - \,
 \frac{m_N}{8\pi^2} \ln\left(\frac{\mu_X}{m_N}\right)
 \mathrm{Re}\big(\bm{h}^\dag \bm{h}\big) \;,
  \label{deltam}
\end{align}
such  that  $\bm{M}_N$ at  the  energy  scale  $m_N$ relevant  for  RL
becomes: $\bm{M}_N(m_N) = m_N \bm{1}_3 + \bm{\Delta M}_N^{\mathrm{RG}}$.
However,  as shown  in~\cite{Dev:2015wpa},  all asymmetries  \emph{per
  lepton flavour} vanish through order $\bm{h}^\dag \bm{h}$.

Let us  briefly review the argument  given in~\cite{Dev:2015wpa}.  The
mass matrix obtained from~\eqref{deltam} is real and symmetric and, as
long as  $|[\Delta M_N^{\rm RG}]_{\alpha  \beta}| \ll m_N$, it  can be
diagonalized  with positive  eigenvalues by  a real  orthogonal matrix
$\bm{O} \in O(3)_{N_R}$:
\begin{equation}\label{eq:diagonalization}
\bm{M}_N(m_N) \ = \ \bm{O} \, \widehat{\bm{M}}_N \, \bm{O}^{T} \;,
\end{equation}
where the caret ( $ \widehat{}  $ ) denotes that the given quantity is
evaluated   in  the   mass  eigenbasis.    Since  $\bm{O}$   is  real,
$\mathrm{Re}(\widehat{\bm{h}}^\dagger  \widehat{\bm{h}})$  is diagonal
too.       Therefore,     all      lepton      flavour     asymmetries
$\varepsilon_{l\alpha}$,                                   proportional
to~\cite{Covi:1996wh,Dev:2014laa}
\begin{align}\label{eq:CPphase}
\varepsilon_{l \alpha} \ &\propto \ {\rm
  Im}\big[\widehat{h}^*_{l\alpha}
  \widehat{h}_{l\beta}(\widehat{h}^\dag
  \widehat{h})_{\alpha\beta}\big]+\frac{m_{N,\,\alpha}}{m_{N,\,\beta}}\:{\rm
  Im}\big[\widehat{h}^*_{l\alpha}
  \widehat{h}_{l\beta}(\widehat{h}^\dag
  \widehat{h})_{\beta\alpha}\big] \notag\\ 
&= \ 2 \, {\rm Im}\big[\widehat{h}^*_{l\alpha}
  \widehat{h}_{l\beta}\big] \, \mathrm{Re}\big[(\widehat{h}^\dag
  \widehat{h})_{\alpha\beta}\big] \; + \; \mathcal{O}(h^6)\;, 
\end{align}
with $\alpha \neq \beta$, vanish through order $h^4$.  Notice that the
inclusion of the second  term in the first line of~\eqref{eq:CPphase},
often omitted in earlier studies, plays a decisive role to obtain this
result. In  order to  circumvent this no-go  theorem, we need  to take
into  account the  leading  imaginary contribution  to~\eqref{deltam}.
This  is contained  in the  second-order term  $\propto t^2_N$  in the
Taylor   series  expansion   of   the  RG   evolution  of   $\bm{M}_N$
in~\eqref{deltam}, i.e.
\begin{align}
 \bm{\Delta  M}_N^{\rm RG} \ &\supset \ i\;  t^2_N \frac{m_N}{(16\pi^2)^2} \, 
    \big[\mathrm{Im}(\bm{h}^\dag \bm{h}), \,\mathrm{Re}(\bm{h}^\dag
      \bm{h})\big] , 
  \label{deltamnew}
\end{align}
with  the  commutator  acting  on  flavour  space.   Notice  that  the
so-generated  imaginary term  in~\eqref{deltamnew}  is logarithmically
enhanced   as   compared   to   the  leading-log   higher   loop-order
contributions  to~\eqref{eq:RG}.   In  addition,  we  neglect  the  RG
effects  of the  Yukawa couplings  in~\eqref{deltam}, as  they  do not
produce extra imaginary terms  in $\bm{\Delta M}_N^{\rm RG}$, at least
up to order $h^4$.

With  the  inclusion   of  both  the  contributions  in~\eqref{deltam}
and~\eqref{deltamnew},  the matrix that  diagonalizes $\bm{M}_N(m_N)$,
as  in~\eqref{eq:diagonalization}, is  now complex  and  unitary, thus
allowing  $\mathrm{Re}(\widehat{\bm{h}}^\dagger  \widehat{\bm{h}})$ to
have non-zero  off-diagonal elements in the mass  eigenbasis. Hence, a
non-zero leptonic  asymmetry $\varepsilon_{l\alpha}$ can  be obtained,
at the  expense of an additional  $\mathcal{O}(h^2)$ suppression. This
additional suppression significantly constraints the allowed region of
the  seesaw mass~$m_N$  in the  radiative MFV  framework,  contrary to
previous expectations~\cite{Branco:2006hz,Deppisch:2010fr}. As  we will see below, new
correlated  heavy- and light-neutrino  mass bounds  can be  deduced by
requiring successful generation of the BAU.

\begin{figure}[t]
\vspace{-0.5em}
\includegraphics[width=0.95\columnwidth]{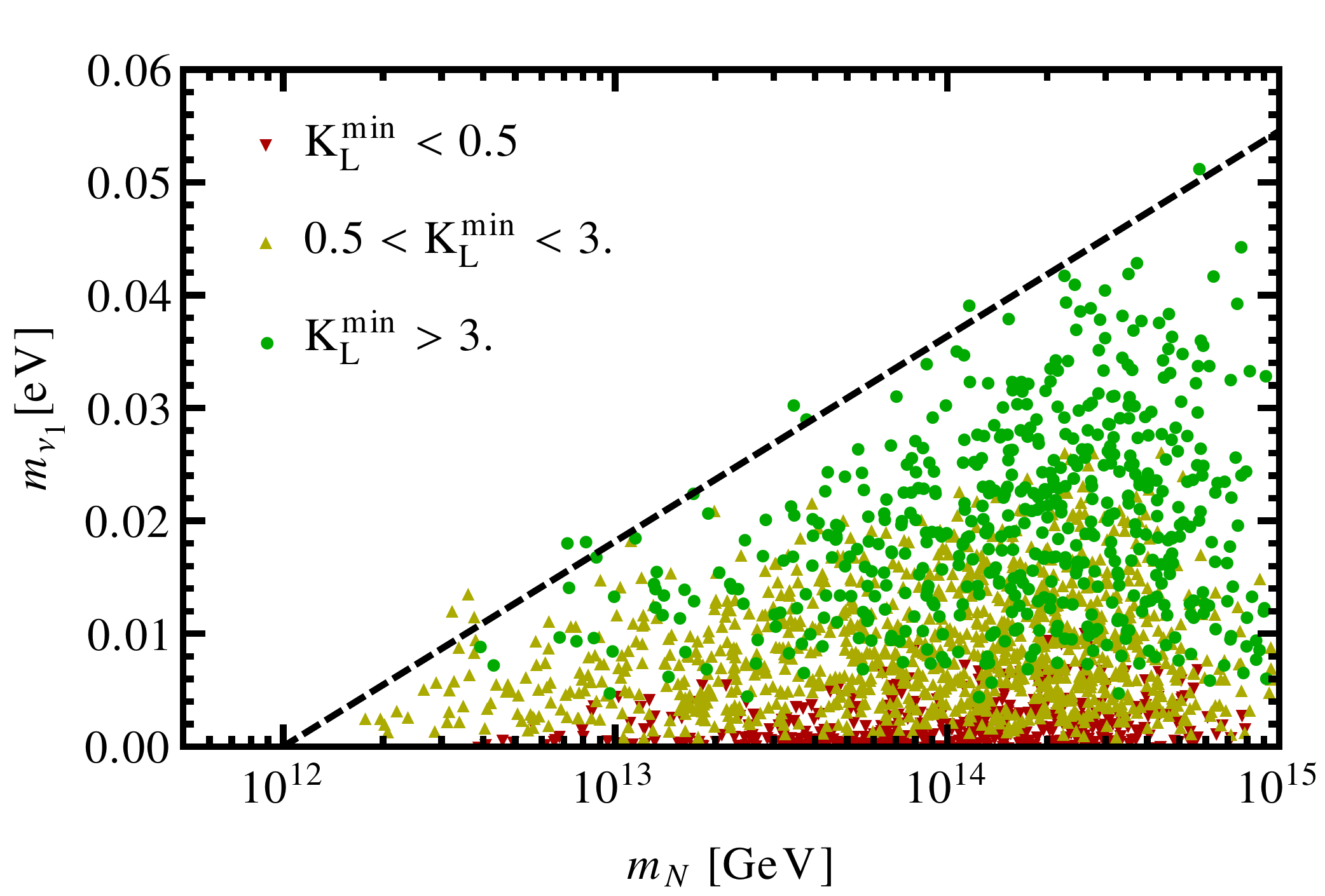}\\
\vspace{-0.2em}
\includegraphics[width=0.95\columnwidth]{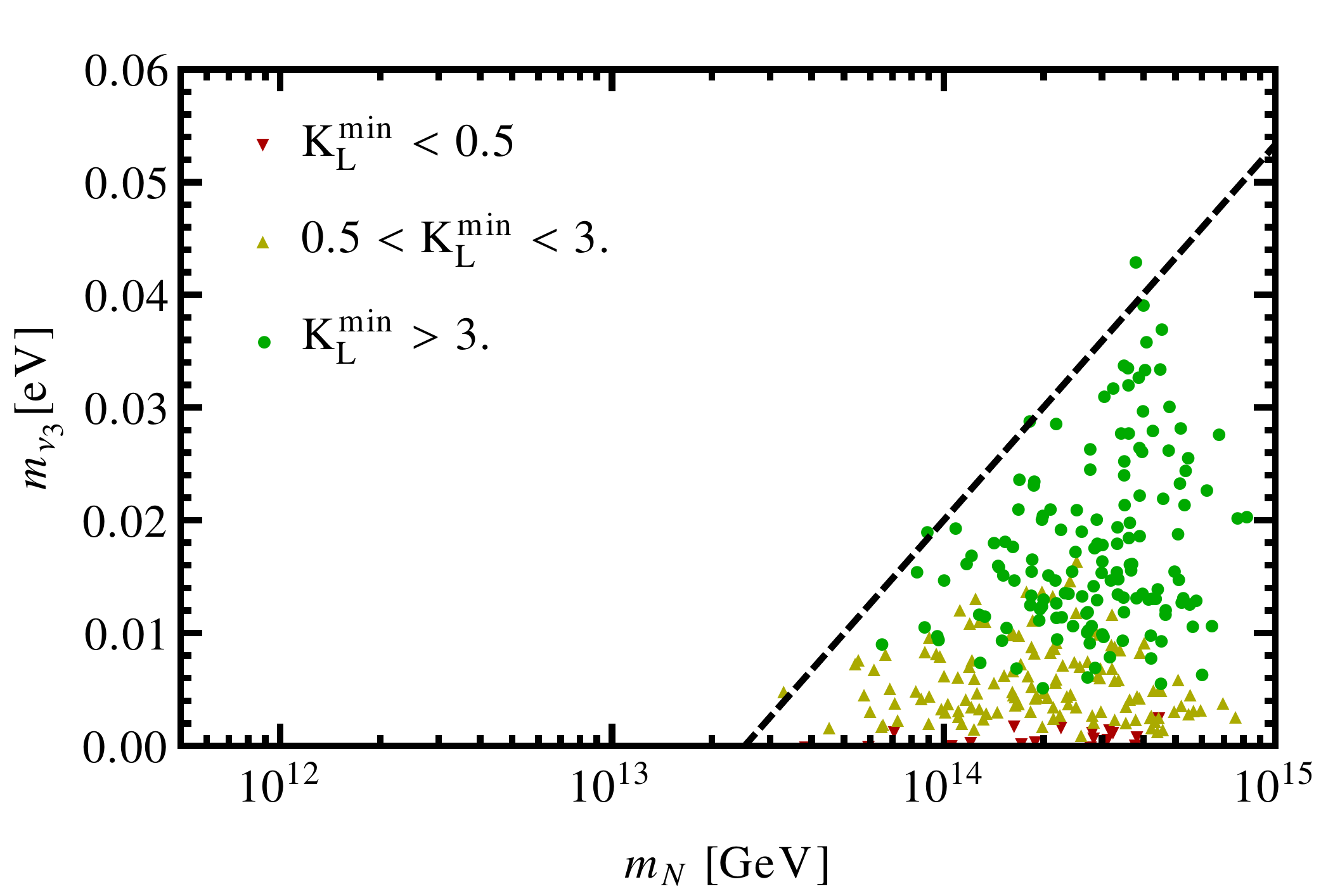}\\
\vspace{-0.4em}
\caption{Scatter plots of the values of the heavy and lightest
  neutrino masses that allow successful leptogenesis, assuming normal
  (top panel) and inverted (bottom panel) hierarchy. 
The  dashed lines denote the approximate bounds given
  in~\eqref{eq:bound}.\label{fig:mass_bounds}}
\end{figure}

In order to  fit the light-neutrino data, we  write $h_{l}^{\phantom l
  \alpha  }$  in  terms  of  an  approximate  parametrization  similar
to~\cite{Casas:2001sr}.   Upon  neglecting  the   heavy-neutrino  mass
splitting in the seesaw  relation, the neutrino Yukawa coupling matrix
$\bm{h}$ may be determined as
\begin{equation}\label{eq:casasibarra}
\bm{h} \ \simeq \ - \, i \, \frac{\sqrt{2 m_N}}{v} \,
\bm{U}_{\mathrm{PMNS}} \, \widehat{\bm{m}}_\nu^{\frac{1}{2}} \,
\bm{R}\;, 
\end{equation}
where  $v=246 \, \mathrm{GeV}$,  $\bm{U}_{\mathrm{PMNS}}$ is  the PMNS
matrix~\cite{Capozzi:2013csa}, $\widehat{\bm{m}}_\nu$  is the diagonal
matrix of  light-neutrino masses and $\bm{R}$ is  a complex orthogonal
matrix that  can be expressed in  terms of three  complex Euler angles
$\psi_{1,2,3}$.   We  then calculate  the  heavy-neutrino mass  matrix
$\bm{M}_N(m_N)$  by  using  the  RG  running  given  in~\eqref{deltam}
and~\eqref{deltamnew}.  We  diagonalize $\bm{M}_N(m_N)$ by  means of a
3-by-3    unitary   matrix~$\bm{U}$   (which    generalizes   $\bm{O}$
in~\eqref{eq:diagonalization})   and   rotate   the  Yukawa   coupling
matrix~$\bm{h}$ to the mass eigenbasis.  We stress here that this last
point,  missed  out in  several  earlier  studies,  becomes of  utmost
importance in  our analysis,  in light of  the no-go  theorem outlined
above.

We calculate  the lepton asymmetry  generated via the RL  mechanism by
means    of   the    flavour-covariant   rate    equations   developed
in~\cite{Dev:2014laa}.  To  simplify our numerical  analysis, we treat
heuristically  the  effect of  \emph{regenerative  $CP$ violation  via
  oscillations   in  medium}~\cite{Dev:2014laa,Dev:2014wsa},   due  to
heavy-neutrino off-diagonal number densities (which quantify the quantum coherences between different flavours), by including a factor of
2 enhancement  with respect to the contribution  of the heavy-neutrino
mixing  alone. The  latter  is obtained  including  only the  diagonal
heavy-neutrino  number  densities  in  the mass  eigenbasis.   In  the
radiative   MFV  scenario  under   study,  this   is  a   rather  good
approximation~\cite{Dev:2014laa},   since  one   is   mostly  in   the
weakly-resonant      regime     in      which     $\Gamma_{N_{\alpha}}
\ll|m_{N_\alpha}-m_{N_\beta}| \ll m_N$, where $\Gamma_{N_{\alpha}}$ is
the width  of $N_{R,\alpha}$. As we will  see, successful leptogenesis
is possible  only for $m_N>10^{12}\,\mathrm{GeV}$ and thus  it is safe
to neglect  the decoherence  effects due to  the SM  Yukawa couplings,
which are  out of equilibrium  in this kinematic region.   However, in
spite  of  this  simplification,  an \emph{unflavoured}  treatment  as
usually  followed  in high-scale  hierarchical  leptogenesis would  be
inappropriate, because  the three  heavy neutrinos decay,  in general,
into three {\em different} superpositions of charged leptons.

\begin{table}
\centering
\begin{tabular}{c|c|c}
\hline
\hline
parameter & \quad lower limit \quad & \quad upper limit \quad\\
\hline\hline
$\log_{10}(m_N/\mathrm{GeV})$ & $11 \, \log_{10}(5)$ & 15\\ 
$\min m_{\nu_l}/\mathrm{eV}$ & 0 & 0.06 \\
$\delta$, $\phi_{1,2}$, $\mathrm{Re}\, \psi_{1,2,3}$ & 0 & $2 \pi$\\
$\mathrm{Im}\, \psi_{1,2,3}$ & -\,3 & 3\\
\hline\hline
\end{tabular}
\caption{Region scanned in the numerical analysis. All the parameters
  are generated randomly, with uniform distributions in the intervals
  quoted.\label{tab:scan}} 
\end{table}

With  these  simplifications,  and  assuming that  the  heavy-neutrino
evolution    is    in    the    strong-washout    regime,    i.e.~that
$\Gamma_{N_\alpha}$   is   much    larger   than   the   Hubble   rate
$H(T\!=\!m_N)$, the  equation for  the lepton-asymmetry due  to mixing
$\bm{\delta  \eta}^L_{\mathrm{mix}}$ can be  written, in  matrix form,
as~\cite{Dev:2014laa}
\begin{equation}\label{eq:rate_eq}
\frac{d}{d z} \, \bm{\delta \eta}^L_{\mathrm{mix}} \ = \ \frac{z^3
  K_1(z)}{2} \, \bigg( \frac{\bm{\varepsilon}}{z} \,-\, \frac{1}{3}
\big\{\bm{\delta \eta}^L_{\mathrm{mix}}, \,\mathbf{K}^{\mathrm{eff}}
\big\}\bigg)\;, 
\end{equation}
where  $z=m_N/T$, with $T$ being the temperature, and $K_1(z)$  is  the  modified  Bessel  function. 
$\bm{\varepsilon}$ and $\mathbf{K}^{\mathrm{eff}}$ are 3-by-3 matrices, in charged-lepton flavour space,
of  leptonic flavour  asymmetries and  K-factors,  respectively. Their
explicit form can be  found in~\cite{Dev:2014laa}. The former describes the CP-asymmetry in the heavy-neutrino decays $N \to L \Phi$ due to flavour mixing, whereas the latter describes the washout of the lepton asymmetry due to inverse decays and scattering processes.

 It is convenient to
rotate~\eqref{eq:rate_eq}       to       the       eigenbasis       of
$\mathbf{K}^{\mathrm{eff}}$,  denoted by a  tilde (  $\widetilde{}$ ),
where    the   evolution    of   the    diagonal    entries   $[\delta
  \widetilde{\eta}^L_{\mathrm{mix}}]_{ll}$    decouples    from    the
respective     off-diagonal    ones.      This    allows     one    to
solve~\eqref{eq:rate_eq} semi-analytically, assuming vanishing initial
asymmetry, obtaining the final asymmetry $\delta \eta^L$ as
\begin{equation}
\delta \eta^L \ \approx \ 2 \times \delta \eta^L _{\mathrm{mix}} \ =
\ 2 \,  \sum_l \, \widetilde{\varepsilon}_{ll} \;
\kappa_{\mathrm{fin}}(\widetilde{\mathrm{K}}^{\mathrm{eff}}_{ll}) \;. 
\end{equation}
The efficiency factor $\kappa_{\mathrm{fin}}$ can be obtained by first 
integrating numerically the differential equation 
\begin{equation}
\frac{d}{d z} \, \kappa \ = \ \frac{z^3 K_1(z)}{2} \, \bigg(
\frac{1}{z} \,-\, \frac{2}{3}
\,\widetilde{\mathrm{K}}^{\mathrm{eff}}_{ll}\,\kappa \bigg)\;,
\end{equation}
and       then       calculating       the      final       stationary
limit~$\kappa_{\mathrm{fin}} = \kappa(z \!\to\! \infty)$.  In  the leptonic strong-washout regime
$\mathrm{K}_L^{\mathrm{min}}          \equiv         \min_l         \,
\widetilde{\mathrm{K}}^{\mathrm{eff}}_{ll}~\gg~1$, the final asymmetry
does  not  depend on  its  initial  value, and  $\kappa_{\mathrm{fin}}
\simeq   3/(2  \widetilde{\mathrm{K}}^{\mathrm{eff}}_{ll})$.   In  the
opposite leptonic  weak-washout limit, one  has $\kappa_{\mathrm{fin}}
\simeq 1$, but a pre-existing asymmetry is not washed out.

\begin{figure}[t]
\vspace{-0.7em}
\includegraphics[width=0.95\columnwidth]{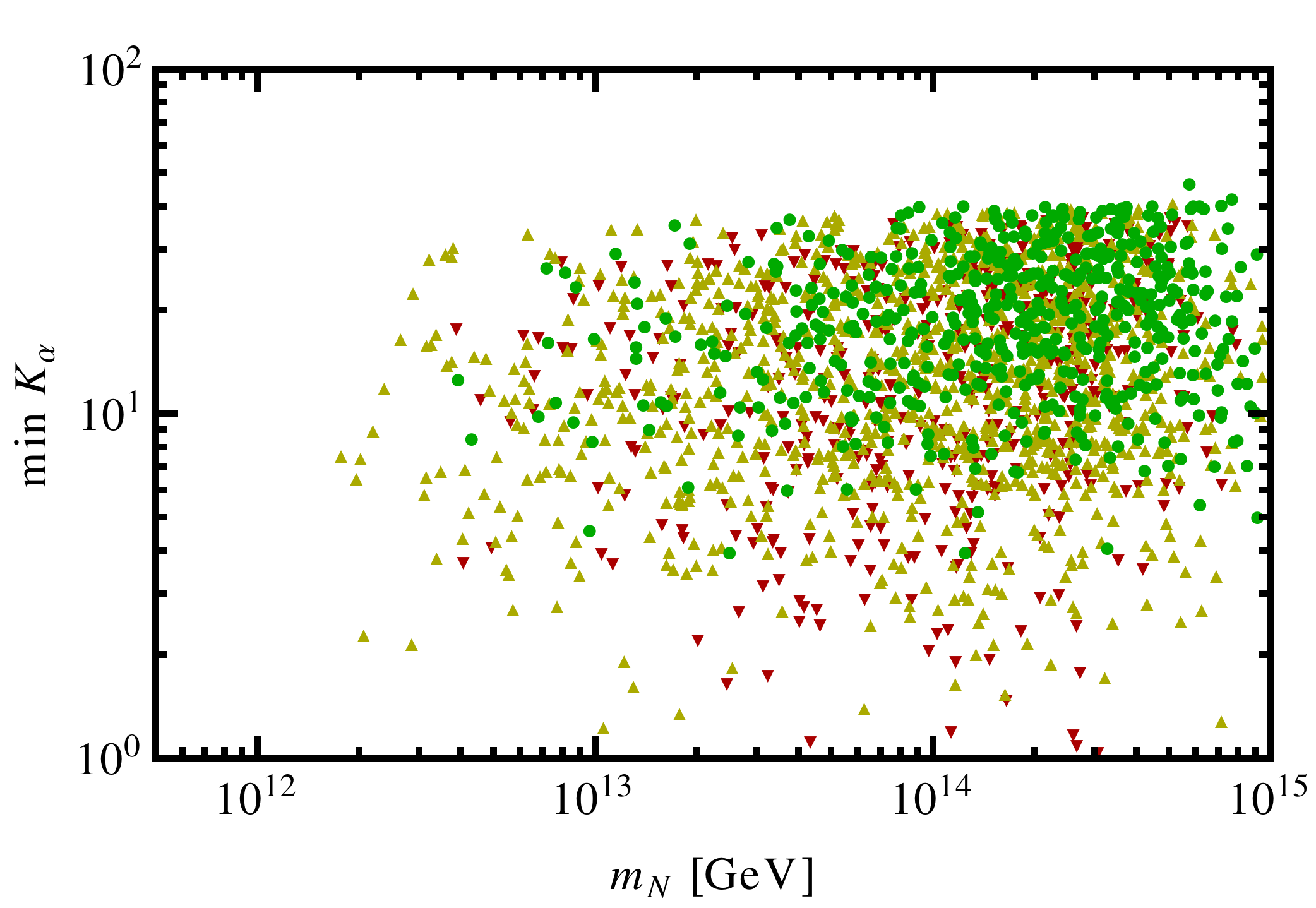}\\
\vspace{-0.4em}
\includegraphics[width=0.95\columnwidth]{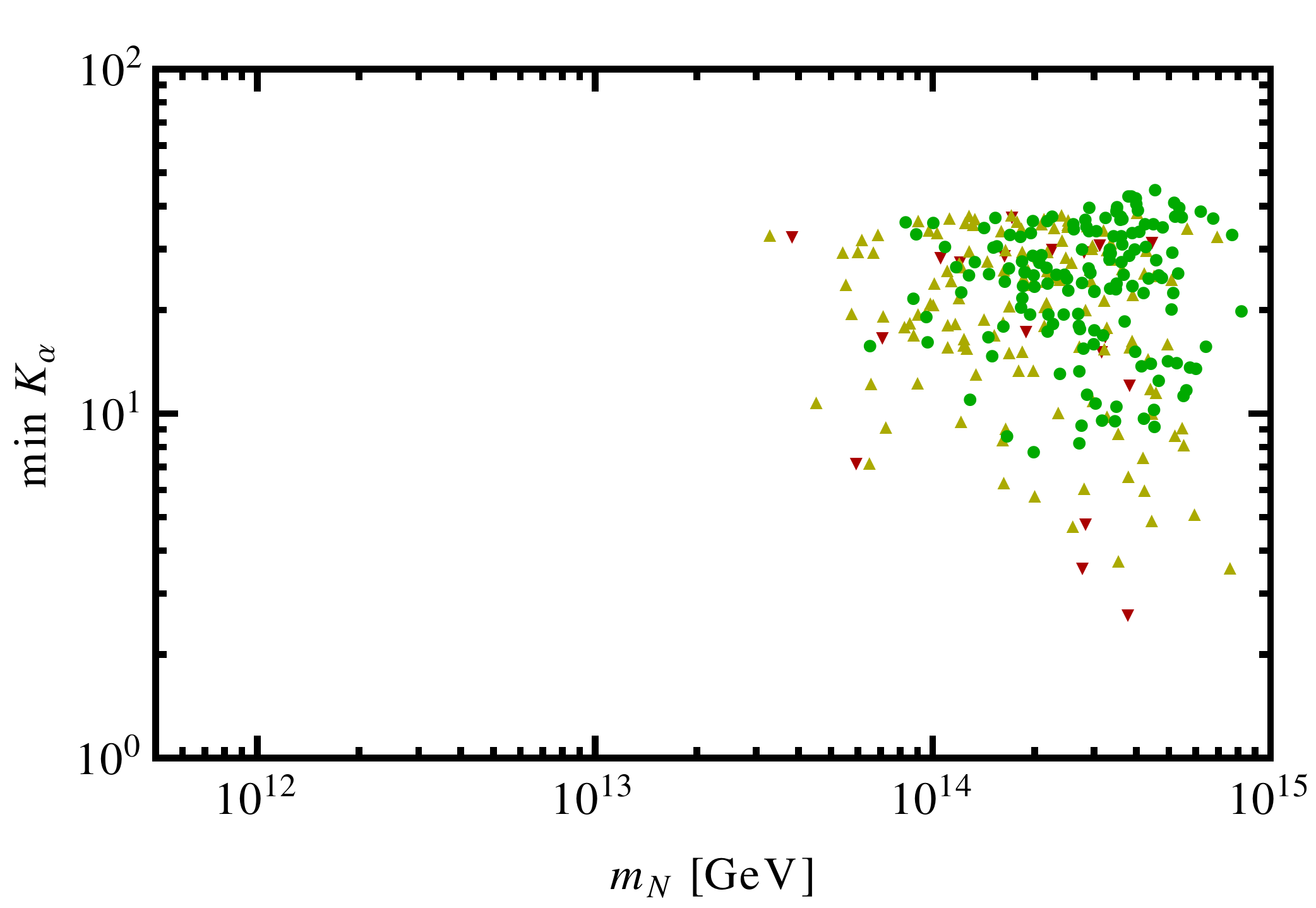}\\
\vspace{-0.5em}
\caption{Smallest heavy-neutrino K-factor, as function of their mass,
  for the numerical scan with normal (top panel) and inverted (bottom
  panel) hierarchy. Legend as in
  Fig.~\ref{fig:mass_bounds}.\label{fig:KN}}
\end{figure}

To study  the constraints derived from successful  leptogenesis in the
radiative MFV scenario,  we perform a numerical scan  of the parameter
space. For both  normal (NH) and inverted hierarchy  (IH) of the light
neutrinos, we generate  a set of $2.5 \times  10^5$ random points. The
parameters of the  scan are: the singlet seesaw  mass scale $m_N$, the
mass of the lightest  neutrino $m_{\nu_{1}}$ ($m_{\nu_{3}}$) in the NH
(IH) case,  the Dirac phase $\delta$ and  Majorana phases $\phi_{1,2}$
of  the  PMNS  matrix  and  the complex  Euler  angles  $\psi_{1,2,3}$
appearing in the matrix $\bm{R}$ in~\eqref{eq:casasibarra}. The limits
of the scanned regions  are exhibited in Table~\ref{tab:scan}. We have
checked   that  enlarging   the   scanned  region   does  not   affect
significantly the results presented  below. In addition, we impose the
following      cuts:     (i)     the      perturbativity     condition
$|\widehat{h}_{l\alpha}|<1$  and  (ii) $\widehat{\mathrm{K}}_\alpha  >
1$,          where         $\widehat{\mathrm{K}}_\alpha         \equiv
\Gamma_{N_\alpha}/(\zeta(3)  H(t\!=\!m_N))$   are  the  heavy-neutrino
washout factors.

\begin{figure}[t]
\vspace{-0.7em}
\includegraphics[width=0.95\columnwidth]{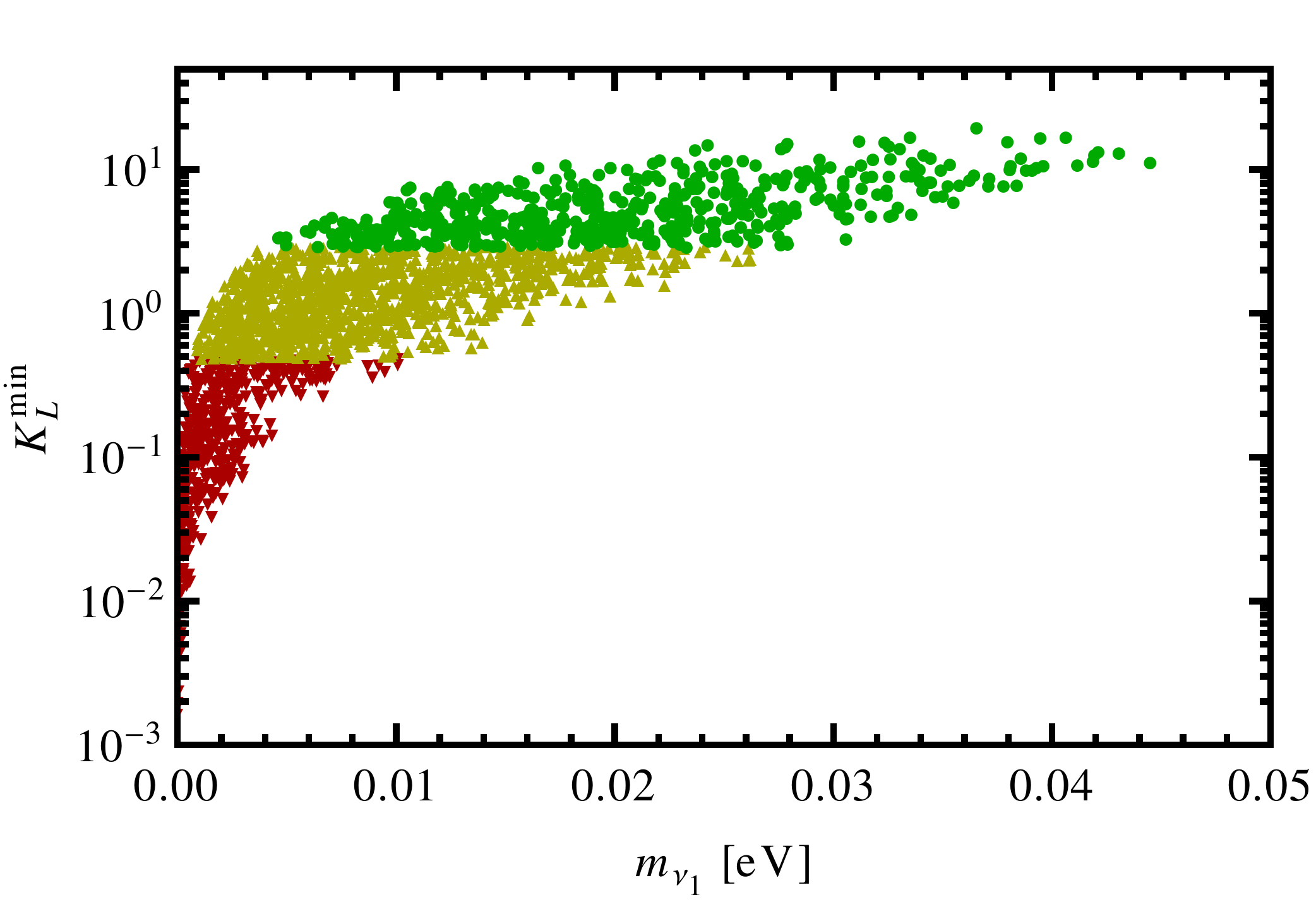}\\
\vspace{-0.7em}
\includegraphics[width=0.95\columnwidth]{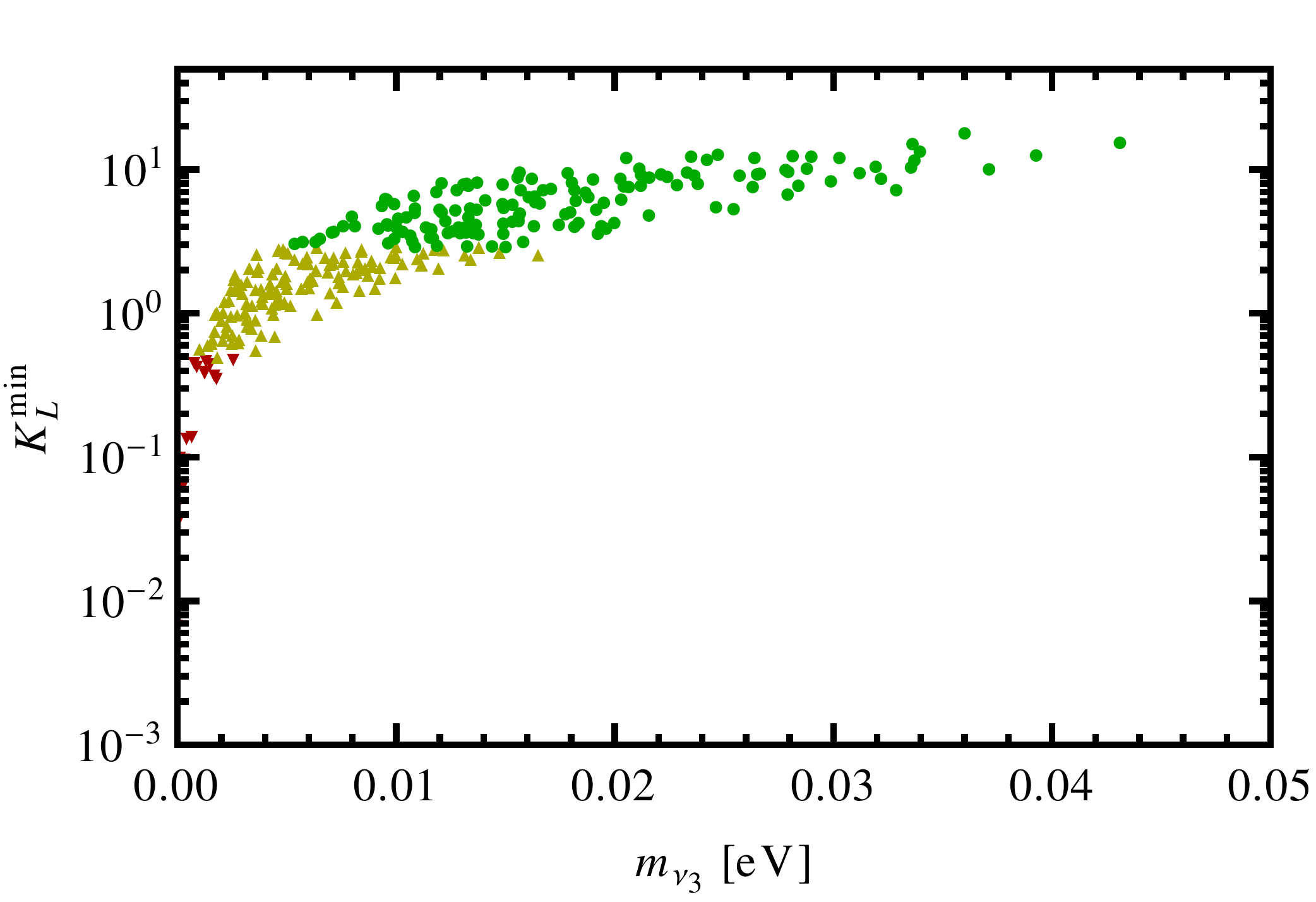}\\
\vspace{-0.7em}
\caption{Smallest effective charged-lepton K-factor $\mathrm{K}_L^{\rm
    min}$, as a function of the lightest-neutrino mass, for the
  numerical scan with normal (top panel) and inverted (bottom panel)
  hierarchy. Legend as in Fig.~\ref{fig:mass_bounds}.\label{fig:KL}}
\end{figure}

In Fig.~\ref{fig:mass_bounds}  we plot the scanned  points that entail
successful     leptogenesis     $|\delta     \eta^L|     >     |\delta
\eta^L_{\mathrm{obs}}|  =  2.47\times 10^{-8}$~\cite{Dev:2014laa},  in
terms  of  the heavy-neutrino  seesaw  mass  $m_N$,  and the  lightest
neutrino masses $m_{\nu_{1,3}}$.   We find that a NH  spectrum for the
light neutrinos  is more pronounced,  with a larger  allowed parameter
space, over  an IH  one.  Moreover, requiring  successful leptogenesis
yields  the following approximate  combined bounds  on the  heavy- and
light-neutrino masses:
\begin{subequations}\label{eq:bound}
\begin{align}
\log_{10} \frac{m_N}{10^{12} \, \mathrm{GeV}} \ &> \ 55 \;
\frac{m_{\nu_1}}{\mathrm{eV}} \;, \quad &\text{(NH)} \\ 
\log_{10} \frac{m_N}{10^{12} \, \mathrm{GeV}} \ &> \ 1.4 \, + \, 30 \;
\frac{m_{\nu_3}}{\mathrm{eV}} \;, \; &\text{(IH)} 
\end{align}
\end{subequations}
In particular, we obtain the approximate absolute bounds:
\begin{equation}\label{eq:bound_abs}
m_N \ > \ 10^{12} \, \mathrm{GeV} \;, \qquad \min m_{\nu_l} \ < \ 0.05
\, \mathrm{eV} \;.
\end{equation}
Observe  that the  light-neutrino  mass bound  is  compatible (at  the
1.9$\,\sigma$  confidence  level)  with  the  recent  evidence  for  a
neutrino mass scale $\sum m_{\nu_l} = (0.320 \pm 0.081)\,\mathrm{eV}$,
obtained from CMB and lensing observations~\cite{Battye:2013xqa}.

In  the rightmost  part  of the  plots in  Fig.~\ref{fig:mass_bounds},
i.e. for  $m_N \sim 10^{15}  \, \mathrm{GeV}$, the  perturbativity cut
$|\widehat{h}_{l\alpha}|<1$  gets  saturated  and  the  error  in  the
approximate   seesaw-inversion  parametrization~\eqref{eq:casasibarra}
reaches  upper values  of $5\%-10  \%$. Also,  in this  region  of the
parameter  space  the  weakly-resonant  condition  is  satisfied  only
mildly. Therefore, we expect our  results to be less accurate for $m_N
\stackrel{>}{{}_\sim}   10^{15}$~GeV.    On   the   other   hand,   in
Fig.~\ref{fig:KN}  we  plot the  successful  points  in  terms of  the
smallest  heavy-neutrino  K-factor.  We  see  that the  heavy-neutrino
strong-washout  requirement  is   comfortably  satisfied  in  all  the
parameter  space.  This  justifies  the simplified  form  of the  rate
equation~\eqref{eq:rate_eq},  for  which  quantum memory  effects  are
omitted.  Finally, in Fig.~\ref{fig:KL}  we show the dependence of the
smallest leptonic  K-factor on the  mass of the lightest  neutrino. We
see  that most  points  have  moderate and  large  values of  leptonic
K-factors in the parameter  space of successful leptogenesis (see also
Fig.~\ref{fig:mass_bounds}). Thus, our predictions for the BAU will be
largely independent  of moderate pre-existing  leptonic asymmetries in
the   MFV  framework.   Requiring   large  leptonic   washout  factors
$\mathrm{K}_L^{\mathrm{min}} >  3$ leads to  an additional approximate
lower  bound   on  the  light   neutrino  masses  $\min   m_{\nu_l}  >
0.004\,\mathrm{eV}$.

The requirement  of successful  MFV leptogenesis restricts  the seesaw
scale $m_N$ of LNV to be  above $10^{12}$~GeV, which is much closer to
the GUT  scale.  It  is therefore natural  to investigate  whether GUT
embeddings of the MFV hypothesis are feasible.  Indeed, a possible MFV
GUT    based     on    $SU(5)$    has     already    been    suggested
in~\cite{Grinstein:2006cg},  but it  would be  interesting  to explore
further this possibility, specifically  within the context of $SO(10)$
unification.   In the  latter case,  a minimal  Higgs content  for the
generation of fermion  masses, with a 10- and  a 126-dimensional Higgs
representations~\cite{Babu:1992ia}, cannot accommodate both degenerate
$N_{R,\alpha}$  and non-trivial  mixing matrices~\cite{Dueck:2013gca}.
Nevertheless, it  appears possible to construct  MFV $SO(10)$ theories
by considering the  full allowed Yukawa sector, which  includes also a
complex  120-dimensional  Higgs   field.   Detailed  studies  of  such
scenarios  are beyond  the  scope of  this  article and  may be  given
elsewhere.

\begin{acknowledgments} The work of A.P. is supported by the 
Lancaster-Manchester-Sheffield Consortium for Fundamental Phy\-sics 
under STFC grant ST/L000520/1.
\end{acknowledgments}

\end{document}